\overfullrule=0pt
\magnification=\magstep1
\baselineskip=5ex
\raggedbottom

\font\sixpoint=cmr6

\def\1{{\bf 1}}
\def\d{{\rm d}}
\def\C{{\bf C}}
\def\R{{\bf R}}
\def\E{{\cal E}}
\def\TF{{\rm TF}}
\def\mfr#1/#2{\hbox{${{#1} \over {#2}}$}}
\def\uprho{\raise1pt\hbox{$\rho$}}

\catcode`@=11
\def\eqalignii#1{\,\vcenter{\openup1\jot \m@th
\ialign{\strut\hfil$\displaystyle{##}$&
        $\displaystyle{{}##}$\hfil&
        $\displaystyle{{}##}$\hfil\crcr#1\crcr}}\,}
\catcode`@=12
\def\boxit#1{\thinspace\hbox{\vrule\vtop{\vbox{\hrule\kern1pt
\hbox{\vphantom{\tt/}\thinspace{\tt#1}\thinspace}}\kern1pt\hrule}\vrule}
\thinspace}

\centerline{\bf QUANTUM DOTS}
\bigskip
\bigskip
\centerline{Elliott H. Lieb\footnote{$^*$}{\sixpoint Work partially
supported by U.S. National Science Foundation grant PHY90-19433 A03} and
Jan Philip Solovej\footnote{$^{**}$}{\sixpoint Work partially supported by
U.S. National Science Foundation grant DMS 92-03829}}
\centerline{\it Department of Mathematics, Fine Hall, Princeton University,
Princeton, NJ  08544}
\bigskip
\centerline{Jakob Yngvason\footnote{$^{***}$}{\sixpoint Work partially 
supported by
the Icelandic Science Foundation.}}
\centerline{\it Science Institute, University of Iceland, Dunhaga 3, IS-107
Reykjavik, Iceland}
\footnote{}{\baselineskip=0.6\baselineskip\hskip -\parindent\sixpoint
To appear in the proceedings of the 
conference on partial differential equations and mathematical physics
held at the University of Alabama, Birmingham, March 1994, International
Press\par
\copyright  1993 in image and
content by the authors. 
Reproduction of this article, by any means, is permitted for non-commercial
purposes.\par}
\bigskip
\bigskip
{\narrower\smallskip\noindent\baselineskip=0.75\baselineskip
{\bf Abstract:}  Atomic-like systems in which electronic motion is two
dimensional are now realizable as ``quantum dots''.  In place of the
attraction of a nucleus there is a confining potential, usually assumed to
be quadratic.  Additionally, a perpendicular magnetic field $B$ may be
present. We review some recent rigorous results for these systems.
We have shown that a Thomas-Fermi type theory for the ground state is
asymptotically correct when $N$ and $B$ tend to infinity.  There are
several mathematically and physically novel features.  1.  The derivation
of the appropriate Lieb-Thirring inequality requires some added effort. 
2.  When $B$ is appropriately large the TF ``kinetic energy'' term
disappears and a peculiar ``classical'' continuum electrostatic theory
emerges.  This is a two dimensional problem, but with a three dimensional
Coulomb potential.  3.  Corresponding to this continuum theory is a
discrete ``classical'' electrostatic theory.  
The former provides an upper bound
and the latter a lower bound to the true quantum energy; the problem of
relating the two classical energies offers an amusing exercise in
electrostatics. \smallskip}
\vfil\eject
\bigskip\noindent
{\bf I.  INTRODUCTION}

Recent developments in semiconductor technology have produced 
remarkably small pieces of matter that behave like two-dimensional atoms. 
These ``dots'' are part of the world of mesoscopic physics which contains
objects that are large compared to atoms but small enough to demonstrate
typical quantum effects such as discrete energy levels  and interference
phenomena.

The simplest quantum dot consists of an insulator, e.g. AlGaAs, and a
semiconductor, e.g. GaAs. Between these two materials is a potential
difference that confines injected electrons to a thin layer at the
interface. The thickness of the layer ranges from 30 to 100 angstroms while
the diameter of the dot, typically of the order 300 to 8000 angstroms, is
controlled by a metal gate on top of the insulator, or by edging a pattern
through the insulator and the interface.
(Recall that $a_{{\rm Bohr}} := \hbar^2/(me^2)$, the radius of a hydrogen
atom, is about half an angstrom, i.e. $0.5\times 10^{-8}$ cm).  The number of
injected electrons, $N$ can conveniently be from zero to several thousand.

Perpendicular to the dot a uniform magnetic field, $B$, can be introduced,
thereby making it possible to investigate the quantum mechanics of motion
in a magnetic field.  The quantized Landau levels can then become prominent.

Although this system is very quantum mechanical, its parameters are
sufficiently different from ordinary atoms and molecules that it is
possible to demonstrate in the laboratory effects that would otherwise
require the magnetic field of a neutron star ($10^9 - 10^{12}$ gauss).
\item{$\bullet$}  Energies are typically in units of
meV (millielectron
volts), not eV.
\item{$\bullet$}  Large magnetic fields need only be several teslas (one
tesla = 10$^4$ gauss).  By a ``large'' field is meant one for which the
effective Landau radius 
is comparable to the effective Bohr radius or smaller.
\item{$\bullet$}  The system is effectively two-dimensional because the
electronic motion in the direction perpendicular to the sandwich is
essentially ``frozen out''.  The thickness, $L_\bot$, is so small that each
electron remains in the ground state, $\phi (x_\bot) = (\sqrt 2/L_\bot)
\sin
(\pi x_\bot/L_\bot)$.  It costs a large kinetic energy to promote an
electron to the next state \hfill\break
$(\sqrt 2/L_\bot) \sin (2 \pi x_\bot /L_\bot)$. 

Whether or not such transitions can be totally disregarded remains to be
proved, but this is the customary assumption and we shall make it here. 
{F}rom the mathematical point of view, the assumption of two-dimensional
motion is the most interesting feature of this model.  Some of our results
here contrast markedly with those in our earlier work on conventional atoms
in large magnetic fields [LSY1-3]. In this paper we present some
recent work; details will appear elsewhere.

Our $N$-body system can be modeled by the following Hamiltonian, $H_N$.
$$H_N = \sum \limits^N_{j=1} H^1_j + e_*^2\sum \limits_{1 \leq i < j \leq
N} \vert
x_i - x_j \vert^{-1}, \eqno(1.1)$$
with $x_i \in \R^2$ and where $H^1$ is the one-body Hamiltonian
$$H^1 = {\hbar^2 \over 2m_*} \left( i \nabla - {e \over \hbar c} A 
\right)^2 + g_* \left( {\hbar e \over 2m_*c} \right) S^z B - \left( {\hbar
e
\over 2m_* c} \right) \left(1 - {\vert g_*\vert \over 2} \right) B + V (x),
\eqno(1.2)$$
with the following notation.

The magnetic vector potential is 
$A = \mfr1/2 B(-y, x)$.

The potential $V(x)$ is confining, which is to say that $V(x) \rightarrow
\infty$ as $\vert x \vert \rightarrow \infty$, and it is assumed to be
continuous.  
For convenience in computations it is usually taken to be of the form $V (x) = 
\mfr1/2 m_*\omega^2 x^2$,
but this is somewhat phenomenological and arbitrary.  It is
created by an external gate and it is therefore adjustable to a certain 
extent.  This $V$
replaces the Coulomb attraction to the nucleus, $-Z /\vert x \vert$, in a
conventional atom.
We write the potential generally as
$$V(x)=Kv(x)$$
and let the coupling constant $K$ and the magnetic field be $N$ dependent,
$$K = \kappa N \ \ {\sl and} \ \ B = \beta N$$
with $\beta$ and $\kappa$ fixed.  In this way the ground state energy,
$E^Q$, 
will scale as $N^2$ as $N \rightarrow \infty$.  (This contrasts with
$N^{7/3}$
for a natural atom.)  

The operator $S^z$ is the perpendicular component of the
spin and takes the values $\pm 1/2$ for each electron.  It commutes with
$H$.

The constant term in (1.2), $- (\hbar e/(2m_*c)) (1-\vert g_*\vert/2) B$,
is included in
order that the ``kinetic energy'' operator, $H^1 - V (x)$, has a spectrum
starting at zero.

The Hilbert space is that appropriate for fermions, the antisymmetric
tensor product $\bigwedge \limits^N_1 L^2 (\R^2; \C^2)$.

The numbers $m_*, e_*, g_*$ are ``effective'' values that differ from the
usual ones because of the electron's interaction with the crystal in which
it finds itself.  The values for GaAs are

$m_* =$ effective mass $= 0.07\ m_{{\rm electron}}$

$e_* =$ effective charge $= e \cdot$ (dielectric constant)$^{-1/2} = 0.3\ e$

$g_* = - .03$

\noindent
Corresponding to these values there are respectively a natural length, 
energy and magnetic field strength:

$a_* = \hbar^2/m_* e^2_* = 185\ a_{{\rm Bohr}} = 10^{-6}$ cm

$E_* = e^4_* m_*/\hbar^2 = 9 \times 10^{-4}\,{\rm Ry} = 
12\ {\rm meV}$

$B_* = e^4_* m^2_* c/e\hbar^3 = 6.7\times 10^4\, {\rm G} = 6.7\ {\rm T}$

\noindent Henceforth, we choose units such that $\hbar=e_*=2m_*=1$. The
energy unit 
is then $2E_*$, and if we moreover measure the magnetic field in units of 
$4B_*$, the numerical factors in (1.1) and (1.2), except 
$g_*$, can be dropped. 
The eigenvalues of $(i \nabla-A)^2$ are the Landau levels $B, 3B, 5B$ in
these units.  The
$g$-factor for normal electrons is 2, which is what appears in the Pauli
operator, and
which leads to eigenvalues 
$0, 2B, 4B,
6B, \dots$ of $(i \nabla- A)^2 + 2S^z B$, with the first being singly
degenerate and the others doubly
degenerate (as far as spin is concerned).  In our case $g_* \not= 2$.  In
fact, it is slightly negative in GaAs!  This merely makes the bookkeeping a
bit more
complicated, but has no essential effect. A derivation of the effective
$g$-factor in a solid 
is given in [KC, pp.\ 282-286].

We draw attention to the fact that the Coulomb repulsion in (1.1) is the
{\it
three-dimensional} potential, $\vert x \vert^{-1}$, even though our system
is two-dimensional.  The absence of Newton's theorem makes our job a little
harder.

If the repulsion is omitted, and if $V$ is quadratic, 
$V(x) = 
\mfr1/4\omega^2 x^2$, the problem is 
solvable. 
The spectrum of $H^1$ was determined by Fock [FV] in 1928, two
years before Landau's paper.  For 
$(i \nabla - A)^2 + \mfr1/4\omega^2 x^2$ the spectrum is given by
$$E = (n_1 - n_2) B + (n_1 + n_2 + 1) [\omega^2 + B^2]^{1/2}
\eqno(1.3)$$
with $n_1, n_2 = 0, 1, 2, \dots$.  
It is remarkable that the simple spectrum, (1.3), gives a qualitatively
good fit to some of the data [KM, KLS].

One of the most important physically measurable quantities is the spectrum
of $H$.  Here we concentrate on the ground state energy $E^Q$ and will
demonstrate that the Thomas-Fermi (TF) energy, which scales precisely as
$N^2$,
is asymptotically exact as $N \rightarrow \infty$ with $\kappa$ and $\beta$
fixed.  The $N \rightarrow
\infty$ limit of $E^Q$ turns out to be {\it uniform} in $\beta$, i.e. given
any $\varepsilon > 0$ we can find $N$ large enough, but independent of
$\beta$, so that $N^{-2} \vert E^{\TF} - E^Q \vert < \varepsilon$.
%
%
\bigskip
\vbox{ \noindent 
{\bf II.  THOMAS-FERMI THEORY}

TF theory is defined by the following functional, $\E^{\TF}$, of a
nonnegative ``density'' function $\uprho = \R^2 \rightarrow [0, \infty)$.}
$$\E^{\TF} (\uprho) = \int \limits_{\R^2} j^{{\phantom{*}}}_B (\uprho (x)) 
\d x + \int
\limits_{\R^2} V(x) \uprho(x) \d x + D (\uprho, \uprho) \eqno(2.1)$$
with
$$D(f,g) = \mfr1/2 \int \limits_{\R^2} \int \limits_{\R^2} f(x) g(y) \vert
x-y \vert^{-1} \d x \d y. \eqno(2.2)$$
The function $j^{{\phantom{*}}}_B (\uprho)$ imitates the kinetic 
energy, $(i \nabla - A)^2 + g_* S^zB - (1 - \vert g_* \vert /2)B$.  It 
is the smallest kinetic energy density for electrons in a
2-dimensional box with particle density $\uprho$.  
For simplicity we give explicit formulas for spinless fermions.
For electrons (with spin) this would correspond to having $g_*=0$ (which
is close to the value for GaAs) except 
for the trivial fact that one must replace $j^{{\phantom{*}}}_B(\uprho)$
by $2j^{{\phantom{*}}}_B (\uprho/2)$ due to the spin degeneracy.
Everything goes through, however,  with obvious 
modifications, for electrons with $g_* \not= 0$.  
If spin is ignored we have $j^{{\phantom{*}}}_B (0) = 0$ and
$j^{{\phantom{*}}}_B (\uprho)$ has slope $2nB$ for $n(B /2 \pi)
\leq \uprho \leq (n+1) (B/2\pi)$, with
$n = 0,1,2, \dots$.  See the graph.  
\bigskip
\hbox{{\vbox{\input epsf
\epsfxsize 0.41\hsize\epsfbox{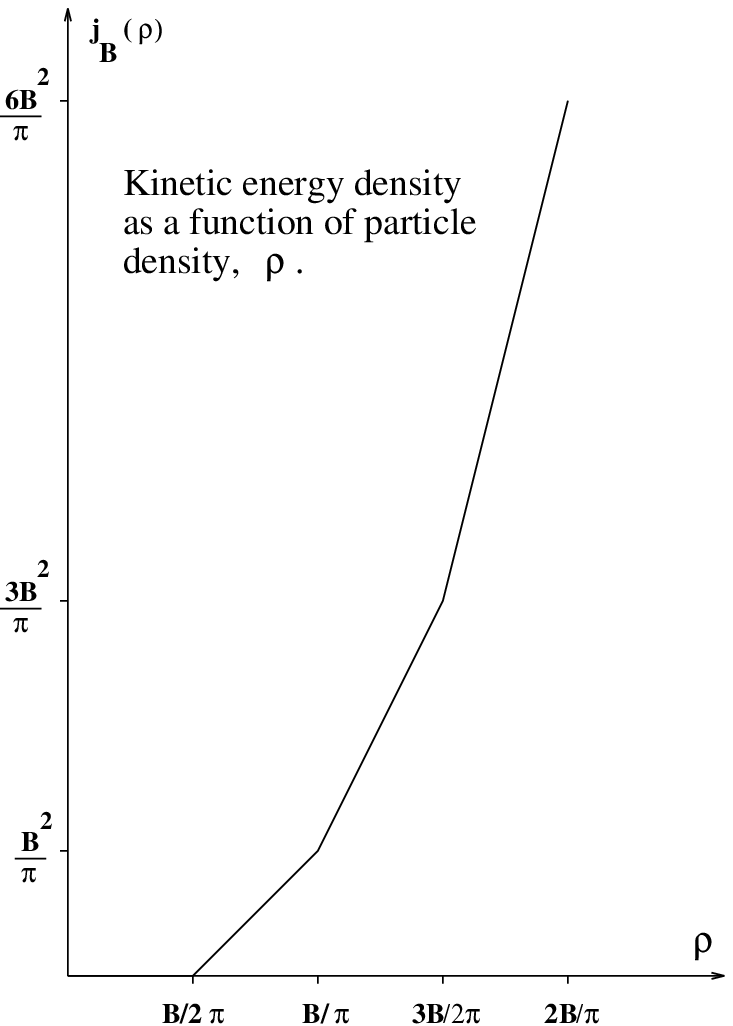}}\hskip 0.10\hsize
\vbox{\hsize=0.451\hsize
The function $j^{{\phantom{*}}}_B$ satisfies $j^{{\phantom{*}}}_B(\uprho)=
B^2j^{{\phantom{*}}}_1(\uprho/B)$. It has
the upper bound \break\hbox{$j^{{\phantom{*}}}_B
(\uprho) \leq 2\pi \uprho^2$}, and $j_B(\uprho)\to j_0 (\uprho) :=
2 \pi \uprho^2$ as $B\to 0$.  The functional $\E^{\TF}$ was first stated by
McEuen {\it et al} [M]
but it was not analyzed completely and it was not proved to give the
asymp\-totic limit of $E^Q$ as $N \rightarrow \infty$.  They call it the
self-consistent (SC) model.  Actually their repulsion term is slightly
different from our $D(\uprho, \uprho)$. 
They replace the Coulomb repulsion
$\vert x-y \vert^{-1}$ by}}}
$$W(x-y) := [\vert x-y \vert^2 + \delta z^2]^{1/2} - [\vert x-y \vert^2 +
4d^2]^{1/2}, \eqno(2.3)$$
where $\delta z \sim 100$ angstroms is the thickness of the dot and $d \sim
1000$ angstroms is the distance to an external gate.  This $W$ has a
short-distance and a long-distance cutoff.  It is, however, positive
definite as a kernel.  Our methods can easily be adapted to prove the
analogues of Theorems 1 and 2 with $\vert x-y \vert^{-1}$ replaced by $W$
everywhere.

The TF energy is defined with $V(x)=Kv(x)$, $K = \kappa N$ and $B = \beta
N$ by
\bigskip
$$E^{\TF} (\kappa, \beta, N) = \inf \biggl\{ \E^{\TF} (\uprho) : \uprho (x)
\geq 0, \int \uprho = N, K = \kappa N, B = \beta N \biggr\} \eqno(2.3)$$
and it is evident that
$$E^{\TF} (\kappa, \beta, N) = N^2 e^{\TF} (\kappa, \beta) \ \ {\sl and} \
\ \uprho^{\TF} (x) = N \uprho^{\TF}_1 (x) \eqno(2.4)$$
where $\uprho_1$ and $e^{\TF}$ are $\uprho$ and $E^{\TF}$ in the special
case $N = 1$.

The functional $\E^{\TF}$ is convex in $\uprho$ and it is strictly convex
since the kernel $\vert x-y \vert^{-1}$ is strictly convex.  Standard
methods yield

{\bf 1.  THEOREM.}  {\it For each $N \geq 0$ and any $g_*$ 
there is a unique minimizer,
i.e., a $\uprho$ such that $\E^{\TF} (\uprho) = N^2 e^{\TF} (\kappa,
\beta)$ and $\int \uprho = N$.}

The only novel feature here is the existence of a minimizer for {\it all}
$N$, which is a consequence of the fact that $V(x) \rightarrow \infty$ as
$\vert x \vert \rightarrow \infty$ instead of $V(x) \rightarrow 0$ as
$\vert x \vert \rightarrow \infty$.

The minimizer satisfies a TF equation, which we illustrate for the 
particular $j_B$ 
defined above.  
Since $j^{{\phantom{*}}}_B
(\uprho)$ is not differentiable at the points $\uprho = n(B/2\pi)$, 
we first have
to define the {\it set valued} function for $\uprho \geq 0$ and $n \geq 0$
given by
$$\eqalignii{j^\prime_B (\uprho) &= \{ 2nB \} \qquad &{\rm if} \quad
n(B/2\pi) < \uprho < (n+1) (B/2\pi) \cr
j^\prime_B (\uprho) &= [2nB, (2n+2)/B] \qquad &{\rm if} \quad \uprho =
(n+1) 
(B/2\pi) \cr} \eqno(2.5)$$
where $\{ x \}$ denotes a single point $x$ and $[a,b]$ is the closed
interval. 
The TF equation can now be written as
$$-Kv(x) - \vert x \vert^{-1} * \uprho + \mu \cases{\in j^\prime_B (\uprho
(x)) &if $\uprho (x) > 0$ \cr
\leq 0 &if $\uprho (x) = 0$ \cr}. \eqno(2.6)$$
This holds also for arbitrary $g_*$ with a suitably modified $j^\prime_B$. 
The quantity $\mu$ is the physical chemical potential, i.e.
$$\mu = {\partial E (K,B,N) \over \partial N} \eqno(2.7)$$
with $K$ and $B$ fixed.

Not only is it true that $\uprho$ satisfies (2.6), but it is also true that
{\it any} (nonnegative) solution pair $(\mu, \uprho)$ to (2.6) is a
minimizing pair, i.e. $\uprho$ is necessarily a minimizer for $\E^{\TF}$
for some $N$ and $\mu$ is the chemical potential (2.7) at that $N$. 
Previously, such an assertion was proved [LS] for 
$j$ a continuously
differentiable function.  The proof when $j^\prime$ is not $C^1$ is a bit
trickier and was carried out by Lieb and Loss [LL].

Our main theorem is the following

{\bf 2.  THEOREM (Convergence of energy and density).}  {\it Fix $\kappa,
\beta$ and $v(x)$.  Then the quantum and TF energies converge in the sense
that
$$N^{-2} \vert E^{\TF} (\kappa, \beta, N) - E^Q (\kappa, \beta, N) \vert
\rightarrow 0$$
as $N \rightarrow \infty$.  The convergence is uniform in $\beta$.  
The densities also converge as $N \rightarrow \infty$:
$$N^{-1} \uprho^Q (x) \rightarrow \uprho^{\TF}_1 (x)$$
in the weak $L^1$ sense.}

Customary techniques, such as coherent states based on Landau orbitals
[LSY2,3], are employed to prove this theorem, but they are insufficient
for the following important reason.  
The function
$j^{{\phantom{*}}}_B$ in (2.1)
is zero in the interval $0 \leq \uprho \leq B/2\pi$.  Correspondingly, the
quantum-mechanical kinetic energy operator $(i \nabla - A)^2 + g_* S^z B -
(1- \vert g_*\vert/2)B$ has
a whole {\it band} of zero modes.  (Note:  It makes no difference if $g_*
\not=
0$ because that will merely entail a shift of the chemical potential.) 
These zero modes cause several problems which we turn to in Sect.~III.

Before doing so, however, let us comment on several unusual features of the
TF density $\uprho (x)$.  For one thing, it always has compact support. 
This is perhaps not surprising since the usual atomic TF density has
compact support except when $N = Z$.  A more interesting feature is that
when $B$ is large, 
$0 < \uprho (x) < B/2\pi$ for all $x$ and hence
$j^{{\phantom{*}}}_B (\uprho (x)) \equiv 0$.  This is not surprising.  When
$B$ decreases, however, $\uprho (x)$ develops a flat spot because (2.6)
causes $\uprho (x)$ to be anchored at 
$B/2\pi$ for a set of positive measure in
$\R^2$.  
The first situation for $v(x) = x^2$, $\kappa = 10^{-4}$ and
$B=8$ tesla (with $N=50$, so $\beta=6\times 10^{-3}$ in our units) is 
shown in Fig.\ 1.
The upper part of the picture shows the density and the lower part
the effective potential $V_{{\rm eff}} = V + \vert x \vert^{-1} * \uprho$
(lower part) are shown.  The
second situation ($B =$ 7 tesla) is shown in Fig.\ 2.  As $B$ is decreased
further the second Landau level starts to be occupied and the flat spot,
caused by the anchoring at $B$, moves further out (Fig.\ 3, $B=5$ tesla). 
For a
still lower field two flat spots appear because 
$2\pi\uprho$ is anchored at $B$ 
and at $2B$.  Eventually more and more flat spots appear.  Fig. 4 shows
three flat spots at 2 tesla.  From
the point of view of the Schr\"odinger kinetic energy, we are seeing the
filling of successive Landau bands.  Also shown on these figures, for
comparison, is a dotted curve representing the solution to a Hartree
calculation of the quantum density for $N=50$.  The curves for the TF 
theory were computed by Kristinn Johnsen while the Hartree calculations 
are due to Vidar Gudmundsson (see [PGM] for details about Hartree
calculations for quantum dots).
\bigskip\noindent
\vfill\eject
{\bf III.  TWO-DIMENSIONAL LIEB-THIRRING INEQUALITY WITH A MAGNETIC 
FIELD}\nobreak

Here we consider spinless fermions to simplify the discussion.
Given any normalized fermionic wave function $\psi (x_1, \dots , x_N)$, 
with $x_i \in \R^d$ we can form the density
$$\uprho_\psi (x) = N \int \limits_{\R^{d(N-1)}} \vert \psi (x_1,
x_2, \dots , x_N) \vert^2 \d x_2 \dots \d x_N,
\eqno(3.1)$$
so that $\int \limits_{\R^d} \uprho_\psi (x) \d x = N$.  The Lieb-Thirring
inequality relates $T_\psi$, the kinetic energy of $\psi$ defined by
$$T_\psi = \left\langle \psi \biggl\vert \sum \limits^N_{i=1} (i \nabla -
A)^2 - B \biggr\vert \psi \right\rangle \eqno(3.2)$$
to some integral of
$\uprho_\psi$.  If $B = 0$ the original bound was $T_\psi \geq K_d \int
\uprho_\psi^{(1+2/d)}$, with $K_d$ being a universal constant that is
positive for all $d \geq 1$.

If we omit the $-B$ term in (3.2) then this original inequality continues
to 
hold for all
vector potentials $A$.  If the $-B$ term is retained, however, the
situation is much more delicate because of the possibility of zero modes
(which certainly exist for $d = 2$, for example, when $B(x)$ has nonzero
total flux).  For $d = 3$ we proved [LSY3] that for some $C_1 > 0$
$$T_\psi \geq C_1 \int j^{(3)}_B (\uprho_\psi (x)) \d x \eqno(3.3)$$
where $j^{(3)}_B (\uprho)$ is the function that appears in the $d = 3$ TF
theory with a constant magnetic field.

In two dimensions we would like to have a similar inequality, but the fact
that $j^{{\phantom{*}}}_B (\uprho) = 0$ for $\uprho \leq B/2\pi$ in
two-dimensions causes difficulties.  The following is adequate, however.

{\bf 3. LEMMA ($d=2$ kinetic energy inequality).}  {\it With $T_\psi$ and
$\uprho_\psi$ defined as above, with $B$ constant in space and with $j_B$
the kinetic energy function defined in Sect. II (i.e., the energy per unit
volume of spinless particles at density $\uprho$), there is a constant
$C_\varepsilon > 0$, for every $0 < \varepsilon < 1$, such that
$$T_\psi \geq C_\varepsilon \int j_B (\varepsilon \uprho_\psi (x)) \d x.
\eqno(3.4)$$}

Inequality (3.4) is equivalent to an assertion about the negative
eigenvalues $e_1 \leq e_2 \leq \dots < 0$ of $H_B = (i \nabla - A)^2 - B -
U(x)$ for an arbitrary nonnegative potential $U(x)$.  One shows that
$$\sum \limits_i \vert e_i \vert \leq \int f(U(x)) \d x$$
from which it follows that $T_\psi \geq \int w(\uprho_\psi (x)) \d x$
with $w$ being the Legendre transform of $f$. We are able to show that for
every $0 < \lambda < 1$
$$\sum \vert e_i \vert \leq  \lambda^{-1} {B \over 2 \pi}\int U + \mfr3/8
(1-
\lambda)^{-2} \int U^2, \eqno(3.5)$$
and (3.4) follows easily from this.

Our proof uses, as usual, the Birman-Schwinger kernel 
$U^{1/2} [(i \nabla - 
A)^2 - B + E]^{-1} U^{1/2}$.  The hard part is to control the zero modes of
$h = (i \nabla - A)^2 -B$, i.e., the lowest Landau level.  Two-dimensions
is
very different from three where the energy in the $z$-direction precludes
having a macroscopically degenerate level of zero energy states.  We are
able 
to isolate this
lowest level with the aid of Ky Fan's inequality on singular values of sums
of
operators; the effect of this level is to produce the $B \int U$ term in
(3.5).

Technicalities aside, (3.4) presents a nontrivial mathematical problem. 
$T_\psi$ can, indeed, be zero; this is easily accomplished, even for
positive density, by making use of all the lowest Landau level
eigenfunctions.  If we
want to bound $T_\psi$ from below by the integral of a {\it nonnegative},
but nontrivial function of $\uprho$, such as $j^{{\phantom{*}}}_B
(\uprho)$,
we have to be sure that the function is zero for low density.  The function
$j^{{\phantom{*}}}_B$, which is zero for $0 \leq \uprho \leq B/2\pi$,
precisely accomodates the maximum density in the lowest Landau band that is
allowed by the Pauli principle.
\bigskip\noindent
{\bf IV.  EXCHANGE-CORRELATION ENERGY AND THE TWO CLASSICAL ELECTROSTATICS
PROBLEMS}\nobreak

The exchange-correlation energy of a normalized state $\psi$ is defined to
be
$$E_{\rm ex} (\psi) = \left\langle \psi \biggl\vert \sum \limits_{1 \leq i
< j
\leq N} \vert x_i - x_j \vert^{-1} \biggr\vert \psi \right\rangle - D
(\uprho_\psi, \uprho_\psi), \eqno(4.1)$$
with $\uprho_\psi$ defined as in (3.1) (but with a sum over spin states)
and $D(f,g)$ defined as in (3.2).  By imitating the original proof in
[LE1],
i.e., using the ``maximal function'', we can prove

{\bf 4.  LEMMA (Exchange inequality in 2 dimensions).}  {\it There is a
constant $C > 0$ such that
$$E_{\rm ex} (\psi) \geq - C \int_{\R^2} \uprho_\psi (x)^{3/2} dx.
\eqno(4.2)$$}

This inequality is quite adequate for small and moderate $B = \beta N$, for
the simple reason that it can be controlled by 
$\int j_B (\uprho_\psi)$ as
follows. 
$$\eqalignno{\int \uprho^{3/2}_\psi &= \int \limits_{\uprho_\psi \leq
B/\pi}
\uprho_\psi^{3/2} + \int \limits_{\uprho_\psi > B/\pi} \uprho_\psi^{3/2}
\cr
&\leq N \sqrt{B/\pi} + \left\{ \int \limits_{\uprho_\psi > B/\pi} 
\uprho_\psi^2
\right\}^{1/2} \left\{ \int \uprho_\psi \right\}^{1/2} \leq \left\{
\sqrt{{\beta 
\over \pi}} + ({\rm const.}) {\sqrt{T_\psi} \over N} \right\} 
N^{3/2}. \qquad&(4.3)\cr}$$
The last line is $O(N^{3/2})$ in the ground state, which is small compared
to the total energy (which is $O(N^2)$) provided one can prove the
``obvious'' fact that $T_\psi =O(N^2)$; this is easy to do.

For large $\beta$ (4.3) is not optimal because it asserts that the error is
$\beta^{1/2} N^{3/2}$.  While this is $O(N^{-1/2})$ relative to $E^Q$, it
is not uniform in $\beta$ and it is certainly not optimal.  The source of
the difficulty, once again, is the zero modes of the kinetic energy
operator.  We can do better by introducing the {\bf
Classical Particle Energy}, $E^P_K$, defined as follows.  First define the
function on $\R^{2N}$ by
$$\E^P_K (x_1, \dots , x_N) = K \sum \limits^N_{i=1} v(x_i) + \sum
\limits_{1 \leq i < j \leq N} \vert x_i - x_j \vert^{-1}. \eqno(4.4)$$
Then $E^P_K$ is defined by
$$E^P_K (N) = \inf \{ \E^P_K (x_1, \dots , x_N) : x_i \in \R^2 \}.
\eqno(4.5)$$

Corresponding to this particle energy is the {\bf Classical Continuous
Energy}:
$$\E^C_K (\uprho) = K \int \limits_{\R^2} \uprho (x) v(x) \d x + D (\uprho,
\uprho) \eqno(4.6)$$
$$E^C_K (N) = \inf \{ \E^C_K (\uprho) : \uprho (x) \geq 0, \int \uprho = N
\}. \eqno(4.7)$$

The classical continuous problem is the same as the TF problem, but without
the kinetic energy $j^{{\phantom{*}}}_B (\uprho)$ term and hence, in
particular, it does not involve $B$.  As a result of the lack of a $\int
\uprho^2$ control, the minimizer, which one can show always exists (since
$V(x) \rightarrow \infty$ as $\vert x \vert \rightarrow \infty$) and is
unique, might turn out to be a measure!  Since $V$ is continuous, this
makes
sense, but we question whether this actually happens.  (Note:  There are
Hausdorff measures in two-dimensions that have finite energy $D(\uprho,
\uprho)$.)  However, as discussed just before Lemma 7, measures to not
occur when $V \in C^{1, \alpha}$.

The classical continuous problem has the same simple scaling property as
the TF
problem, namely
$$E^C_K (N) = N^2 E^C_\kappa (1) \quad {\sl and} \quad \uprho_N (x) = N
\uprho_1 (x), \eqno(4.8)$$
where $\uprho_N$ is the minimizer (with $\kappa$ fixed) for $\int \uprho =
N$ and $K = \kappa N$.

The minimization problem (4.7) leads, of course, to the following
Euler-Lagrange equation for $\uprho$ that parallels the TF equation (2.6). 
For $N = 1, K = \kappa$
$$V_{{\rm eff}} (x) := \kappa v(x) + [\vert x \vert^{-1} * \uprho] (x) =
\cases{\mu &if $\uprho (x) > 0$ \cr \geq \mu &if $\uprho (x) = 0$ \cr}
\eqno(4.9)$$
with $\mu$ being the chemical potential.  [Note:  if $\uprho$ is a measure
the left side of (4.9) should be $\geq \mu$ everywhere and $= \mu \
\uprho$-almost everywhere.

Eq. (4.9) states that the potential generated by $\uprho$ plus the
confining potential $V$ add up to a constant when $\uprho > 0$.  {\it If}
we were in three-dimensions instead of two, the solution is easy.  Just
apply $-\Delta$ to both sides.  If $v(x) = x^2$ one has that $\uprho$ is a
constant, $C$, inside some ball of radius $R$ (with $C = 3\kappa /(2\pi), \
2 \kappa R^3 = 1, \ \mu = 1/R + \kappa R^2)$.

The solution to (4.9) in two-dimensions is much more difficult.  However,
when $v(x) = x^2$ it can be solved:
$$\uprho (x) = \cases{C \sqrt{R^2 - x^2} &$\vert x \vert \leq R$ \cr
0&$\vert x \vert \geq R$ \cr} \eqno(4.10)$$
for suitable constants $C$ and $R$.  The computation of the convolution
$\vert x \vert^{-1} *$ (right side of (4.10)) is an amusing exercise!

We have three non-quantum energies before us, $E^{\TF} (\kappa, \beta, N),
E^P_K (N), E^C_K (N)$.  In order to discuss their relationship we first
need a technical lemma

{\bf 5.  LEMMA (Finite radius).}  {\it Fix $\kappa$ and set $K = \kappa N$.

There is a radius $R_v (\kappa)$, independent of $N$, such that for every
$N$ the infimum for $E^{\TF}, E^P, E^C$ can be sought among densities (in
the case of $E^{\TF}$ and $E^C$), or particle coordinates (in the case of
$E^P$) that have support in the ball of radius $R_v (\kappa)$ centered at
the origin in $\R^2$.}

The proof of Lemma 5 is fairly simple.  If there is some charge outside
some large ball, $R_v (\kappa)$, then there is some location inside the
ball of unit radius to which this charge can be moved in such a way that
the energy is lowered.  Here, we make use of the fact that $V(x)
\rightarrow \infty$ as $\vert x \vert \rightarrow \infty$.

In the remainder of this section we discuss the relationship between the
two classical energies $E^P_K (N)$ and $E^C_K (N)$.  In the next section we
discuss the relationship of these classical energies to $E^{\TF}$.  We
first point out that Lemma 4 cannot be used in order to compare $E^P_K$ and
$E^C_K$.  In fact, what corresponds to the minimizing ``density''
$\uprho_\psi$ for the $\E^P_K$ problem is a sum of delta functions and thus
$\int \uprho^{3/2}_\psi$ is meaningless (if we approximate the delta
functions by continuous functions this error term will go to infinity). 

Using an electrostatics inequality of Lieb and Yau [LY] we are able to
prove that $E^C_K (N)$ is indeed a good approximation to the particle
energy $E^P_K (N)$.  That electrostatics inequality generalizes (4.2) in a
certain sense.  It is the following

{\bf 6.  LEMMA (The interaction of points and densities).}  {\it 
Given points $x_1, \ldots, x_N$ in ${\bf R}^2$, we define Voronoi
cells $\Gamma_1,\ldots, \Gamma_N\subset {\bf R}^2$ by
$$
	\Gamma_j=\{y\in {\bf R}^2 : |y-x_j|\leq |y-x_k|\ \ 
	\hbox{for all \ } k\ne j\} \quad.
$$
These $\Gamma_j$ have disjoint interiors  and their union covers
${\bf R}^2$. We also define $R_j$ to be the distance from $x_j$
to the boundary of $\Gamma_j$, i.e., $R_j$ is half the distance 
of $x_j$ to its nearest  neighbor. Let $\uprho$ be any (not necessarily
positive) function on ${\bf R}^2$. (In general, $\uprho$ can be replaced
by a measure, but it is not necessary for us to do so.) Then
$$
	\eqalignno{\sum_{1\leq i<j\leq N} |x_i-x_j|^{-1}\geq&
	- D(\uprho, \uprho)
	+ \sum_{j=1}^N\int_{{\bf R}^2} \uprho(y)|y-x_j|^{-1}dy\cr
	&{}+\mfr1/8 \sum_{j=1}^N R_j^{-1} 
	-\sum_{j=1}^N \int_{\Gamma_j}\uprho(y) |y-x_j|^{-1}dy\quad.
	&(4.11)}
$$
}

This comparison of $E^C_K (N)$ and $E^P_K(N)$ requires that the minimizing
density for the continuous energy $\E^C_K$ has some regularity.  A-priori
the minimizing density for $\E^C_K$ {\it could be} a singular measure and
if this is the case we cannot expect to get a good estimate of the
difference 
between $E^C_K$ and $E^P_K$.  To ensure that the minimizer for $\E^C_K$ is
neither a measure not too singular a function we need the confining
potential 
$V$ to be sufficiently
regular.  Although we do not believe this to be optimal, we find it
convenient to assume that $V$ is in $C^{1, \alpha}$ (H\"older continuous
first derivative with exponent $\alpha>0$).  We prove that $\uprho$ is then
in $L^q (\R^2)$ for some exponent $q > 2$ that depends on $\alpha$.  With
this in hand, we are then able to prove

{\bf 7.  LEMMA (Classical correlation estimate).}  {\it Assume that $V$ is
a potential in $C^{1, \alpha}$.  For all $N$ we have
$$E^C_K (N) - b (\kappa) N^{3/2} \leq E^P_K (N) \leq E^C_K (N) - a (\kappa)
N^{3/2}, \eqno(4.12)$$
where $a (\kappa)$ and $b (\kappa)$ are positive constants independent of
$N$ that depend on the radius $R_v (\kappa)$ of Lemma 5 and on the exponent
$q$ (and its dual $p = q/(q-1)$).  
$$\eqalignno{a (\kappa) &= (8 R_v (\kappa))^{-1} \cr
b (\kappa) &= \mfr4/3 \sqrt{\mfr2/3} \int \uprho_1 (x)^{3/2} \d x + \left( 
{2 \pi \over 2-p} \right)^{1/p} (2 R_v (\kappa))^{-1+2/p} \Vert \uprho_1
\Vert_q, \cr}$$
with $\uprho_1$ being the minimizer for $\E^C_K$ for $N = 1$, i.e., the
minimizer for $\E^C_\kappa$.}
\bigskip\noindent
{\bf V.  OUTLINE OF THE PROOF OF THE LIMIT THEOREM}\nobreak

We now give a very sketchy outline of the proof of Theorem 2.  We shall
mainly emphasize the steps in the proof that do not use standard
techniques.  In particular, we shall stress the use of the estimates given
in the two previous sections.  

The aim is to prove that for a given $\varepsilon > 0$ we can find an
electron number $N(\varepsilon, \kappa)$, depending on $\varepsilon$ and
$\kappa$ but {\it independent of} $\beta$, such that $N \geq N
(\varepsilon, \kappa)$ implies
$$N^{-2} \vert E^{\TF} (\kappa, \beta, N) - E^Q (\kappa, \beta, N) \vert <
\varepsilon \eqno(5.1)$$
for all $\beta$.  To prove this we have to treat small and large $\beta$
differently.  We therefore choose $\beta_\varepsilon$ (to be specified
below) and consider $\beta \leq \beta_\varepsilon$ and $\beta >
\beta_\varepsilon$. In both cases the proof consists of finding upper and
lower bounds to $E^Q (\kappa, \beta, N)$.  As usual, the upper bound is a
relatively simple variational argument.  In both cases we construct a trial
wave function using magnetic coherent states similar to the ones used in
[YJ] and [LSY1-3].  The lower bounds are more difficult; the case of large
$\beta$, in particular, requires new ideas---again due to the presence of a
large number of zero modes.  We discuss this large $\beta$ case first.

{\it Case 1.  (Lower bound on $E^Q$ for $\beta > \beta_\varepsilon$):} 
By simply ignoring the kinetic energy operator, which we had normalized to
be positive, we have the obvious inequality $E^Q (\kappa, \beta, N) \geq
E^P_K (N)$. 

{F}rom Lemma 7 we can therefore conclude that
$$E^Q (\kappa, \beta, N) \geq E^C_K (N) - b (\kappa) N^{3/2}. \eqno(5.2)$$

The final step in the proof is to relate $E^C_K (N)$ to $E^{\TF} (\kappa,
\beta, N)$.  In order to do this we choose a {\it bounded} density
$\widetilde{\uprho}_1$ with $\int \widetilde{\uprho}_1 = 1$ such that
$\E^C_\kappa (\widetilde{\uprho}_1) \leq E^C_K (1) + \varepsilon /2$. 
Note that $\E^C_\kappa$ does not depend on $\beta$.  We may then assume
that $\beta_\varepsilon$ is so large that 
$\widetilde{\uprho}_1 (x) \leq
\beta_\varepsilon /2\pi$ for all $x$.  For all $\beta > \beta_\varepsilon$
we therefore have 
$j^{{\phantom{*}}}_{\beta N} (N
\widetilde{\uprho}) = 0$ (for simplicity we are here again ignoring spin). 

Consequently, using $\widetilde{\uprho} = N
\widetilde{\uprho}_1$ as a trial density in the TF functional we find that
$$E^{\TF} (\kappa, \beta, N) \leq \E^{\TF} (\widetilde{\uprho}) = \E^C_K
(\widetilde{\uprho}) = N^2 [E^C_K (1) + \varepsilon /2] \leq E^C_K (N) +
{\varepsilon \over 2} N^2. \eqno(5.3)$$
Comparing (5.2) and (5.3) we immediately see that $N(\varepsilon, \kappa)$
can be chosen so that (5.1) holds.

In the special case of $v(x) = x^2$ we know that the minimizer $\uprho_1$
of
$\E^C_\kappa$ given by (4.10) is continuous and bounded.  In this 
case we can therefore
choose $\widetilde{\uprho}_1 = \uprho_1$ and we can make the explicit
choice
$\beta_\varepsilon = \sup_x 2 \pi \uprho_1 (x) = 2 \pi \uprho_1 (0) = CR$.

{\it Case 2.  (Lower bound on $E^Q$ for $\beta \leq
\beta_\varepsilon$):}  In this case we follow the standard route of using
inequalities (4.2) and (4.3) to reduce the many-body problem to a one-body
problem.  Combining (1.1), (4.1-3) we obtain
$$\langle \psi \vert H_N \vert \psi \rangle = \sum \limits^N_{j=1} \langle
\psi \vert H^1_j \vert \psi \rangle + D (\uprho_\psi, \uprho_\psi) - C
(\kappa, \varepsilon) N^{3/2}, \eqno(5.4)$$
where $C(\kappa, \varepsilon)$ is chosen as an estimate on the factor in
$\{ \ \ \}$ on the right side of (4.3), i.e., $\sqrt{\beta \varepsilon/\pi}
+ ({\rm const.}) \sqrt{T_\psi /N} \leq C(\kappa, \varepsilon)$.

To relate (5.4) to the Thomas-Fermi problem we use the inequality
$$0 \leq D (\uprho_\psi - \uprho^{\TF}, \uprho_\psi - \uprho^{\TF}) = -
\left\langle \psi \biggl\vert \sum \limits^N_{j=1} \uprho^{\TF} * \vert x_j
\vert^{-1} \biggr\vert \psi \right\rangle + D (\uprho_\psi, \uprho_\psi) +
D(\uprho^{\TF}, \uprho^{\TF}),$$
which is a consequence of the positive definiteness of the kernel $\vert
x-y \vert^{-1}$.  Inserting this in (5.4) gives
$$\langle \psi \vert H_N \vert \psi \rangle \geq \sum \limits^N_{j=1}
\left\langle \psi \biggl \vert H^1_j + \sum \limits^N_{j=1} \uprho^{\TF} *
\vert x_j \vert^{-1} \biggr\vert \psi \right\rangle - D (\uprho^{\TF},
\uprho^{\TF}) - C (\kappa, \varepsilon) N^{3/2}. \eqno(5.5)$$
The importance of (5.5) is that the quantum energy, $\langle \psi \vert H_N
\vert \psi \rangle$, is bounded below by the expectation of the one-body
operator $H^1 + \uprho^{\TF} * \vert x \vert^{-1}$.

To prove Theorem 2 it remains to do a semiclassical analysis (with
$N^{-1/2}$ playing the role of an effective Planck's constant) of the sum
of the negative eigenvalues of this one-body operator.  This semiclassical
analysis again uses the magnetic coherent states and is very similar to the
analysis in [LSY3].
\bigskip\noindent
{\bf Acknowledgements} We thank Kristinn Johnsen and Dr.\ Vidar 
Gudmundsson for valuable comments and for allowing us to use 
the pictures of densities and 
effective potentials that they prepared. 
\vfill\eject
{\baselineskip=4ex
{\bf REFERENCES}
\item{[FV]}  V. Fock, {\it Bemerkung zur Quantelung des harmonischen
Oszillators in Magnetfeld}, \hfill\break 
Z. Phys. {\bf 47}, 446-448 (1928).
\item{[KC]}  C. Kittel, {\sl Quantum Theory of Solids}, Wiley (1963).
\item{[KLS]}  A. Kumar, S.E. Laux and F. Stern, {\it Electron states in a
GaAs quantum dot in a magnetic field}, Phys. Rev. B. {\bf 42}, 5166-5175
(1990).
\item{[KM]}  M.A. Kastner, {\it Artificial atoms}, Physics Today {\bf 46},
24-31 (1993).
\item{[LE1]}  E.H. Lieb, {\it A lower bound for Coulomb energies}, Phys.
Lett. {\bf 70A}, 444-446 (1979).
\item{[LL]}  E.H. Lieb and M. Loss, unpublished section of a book on
stability of matter.
\item{[LS]}  E.H. Lieb and B. Simon, {\it Thomas-Fermi theory of atoms,
molecules and solids}, Adv. in Math. {\bf 23}, 22-116 (1977).
\item{[LSY1]}  E.H. Lieb, J.P. Solovej and J. Yngvason, {\it Heavy atoms in
the strong magnetic field of a neutron star}, Phys. Rev. Lett. {\bf 69},
749-752 (1992).
\item{[LSY2]}  E.H. Lieb, J.P. Solovej and J. Yngvason, {\it Asymptotics of
heavy atoms in high magnetic fields:  I.  Lowest Landau band region},
Commun. Pure Appl. Math. (in press).
\item{[LSY3]}  E.H. Lieb, J.P. Solovej and J. Yngvason, {\it Asymptotics of
heavy atoms in high magnetic fields:  II.  Semiclassical regions}, Commun.
Math. Phys. {\bf 161}, 77-124 (1994).
\item{[LY]}  E.H. Lieb and H.-T. Yau, {\it The stability and instability of
relativistic matter}, Commun. Math. Phys. {\bf 118}, 177-213 (1988).
\item{[M]}  P.L. McEuen, E.B. Foxman, J. Kinaret, U. Meirav, M.A. Kastner,
N.S. Wingreen and S.J. Wind, {\it Self consistent addition spectrum of a
Coulomb, island in the quantum Hall regime}, Phys. Rev. B, {\bf 45},
11419-11422 (1992).  Cf. eqn. (3).
\item{[PGM]} D.~Pfannkuche, V.~Gudmundsson, P.A.~Maksym, 
{\it Comparison of a Hartree, a
Hartree-Fock, and an exact treatment of quantum dot helium}, Phys. Rev. B,
{\bf 47}, 2244-2250 (1993).
\item{[YJ]}  J.~Yngvason, {\it Thomas-Fermi theory for matter in a magnetic
field as a limit of quantum mechanics}, Lett. Math. Phys. {\bf 22}, 107-117
(1991).\par}
\vfil\eject
\vbox{\vskip -1truecm\input epsf
\epsfysize 1.1\vsize\epsfbox{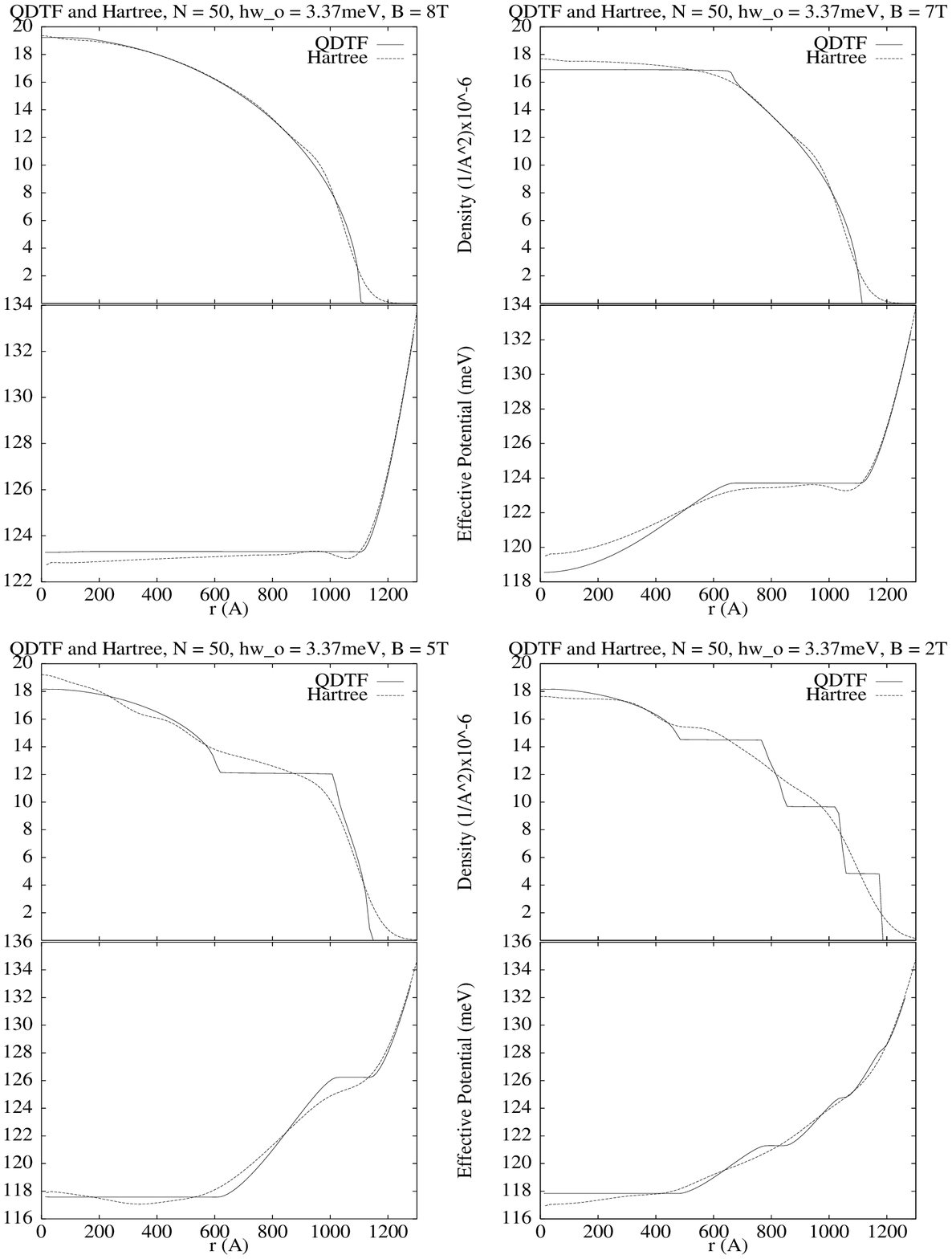}}

\bye